\documentclass[10pt]{iopart}

\usepackage{amssymb}
\usepackage{graphicx}
\usepackage{dcolumn}
\usepackage[colorlinks,urlcolor=blue,citecolor=blue,linkcolor=blue]{hyperref}
\usepackage{lipsum}
\usepackage[dvipsnames]{xcolor} 
\usepackage{iopams}
\usepackage{lineno}
\usepackage{xspace}
\usepackage[url=false, giveninits=true]{biblatex}
\usepackage{amsmath}
\usepackage{subfigure}

\usepackage{siunitx}

\newcommand{\dynesty}{\textsc{dynesty}\xspace}
\newcommand{\bilby}{\textsc{bilby}\xspace}

\newcommand{\nrsur}{\textsc{NRSur7dq4}\xspace}

\newcommand{\imrx}{\textsc{IMRPhenomXHM}\xspace}
\newcommand{\imrxp}{\textsc{IMRPhenomXPHM}\xspace}
\newcommand{\python}{\textsc{Python}\xspace}
\newcommand{\gwmemory}{\textsc{gwmemory}\xspace}

\begin{document}

\title[Searching for gravitational-wave memory in the third LIGO-Virgo-KAGRA gravitational-wave transient catalogue]{Does spacetime have memories? Searching for gravitational-wave memory in the third LIGO-Virgo-KAGRA gravitational-wave transient catalogue}

\author{Shun Yin Cheung}
    \ead{shun.cheung@monash.edu}
\author{Paul D. Lasky}
    \ead{paul.lasky@monash.edu}
\author{Eric Thrane}
    \ead{eric.thrane@monash.edu}

\begin{abstract}
Gravitational-wave memory is a non-linear effect predicted by general relativity that remains undetected.
We apply a Bayesian analysis framework to search for gravitational-wave memory using binary black hole mergers in LIGO-Virgo-KAGRA's third gravitational-wave transient catalogue. 
We obtain a Bayes factor of $\ln \text{BF}=0.01$, in favour of the no-memory hypothesis, which implies that we are unable to measure memory with currently available data.
This is consistent with previous work, suggesting that a catalogue of $\mathcal{O}(2000)$ binary black hole mergers is needed to detect memory. 
We look for new physics by allowing the memory amplitude to deviate from the prediction of general relativity by a multiplicative factor $A$.
We obtain an upper limit of $A<23$ ($95\%$ credibility).
\end{abstract}

\submitto{\CQG}

\maketitle \ioptwocol

\section{Introduction}\label{sec:introduction}

The landmark detection of gravitational waves from the merger of a binary black hole in 2015 by the LIGO-Virgo Scientific Collaboration has provided new methods to test general relativity and fundamental physics \cite{abbott_observation_2016, ligo_scientific_collaboration_and_virgo_collaboration_tests_2019}. However, a particularly interesting phenomenon predicted by general relativity remains unconfirmed: \emph{gravitational-wave memory}. Linear memory was first predicted by Zel'dovich and Polnarev and is produced from unbound systems such as hyperbolic orbits, supernovae and triple black hole interactions \cite{zeldovich_radiation_1974, braginsky_gravitational-wave_1987}. In 1991, Christodoulou identified a significant non-linear memory component in bound systems, such as compact binary mergers \cite{christodoulou_nonlinear_1991}. Non-linear memory arises from the gravitational waves themselves, resulting in an accumulation of memory that physically manifests as a \textit{permanent} displacement between test masses following the passage of gravitational waves \cite{thorne_gravitational-wave_1992}.

The displacement memory signal has not yet been directly observed because the amplitude of the memory is only around $\lesssim5\%$ of the oscillatory waveform amplitude for a GW150914-like event \cite{lasky_detecting_2016}. Due to the low amplitude of memory, a direct detection of memory from a single event is improbable with current gravitational-wave detectors, unless observatories detect a surprisingly close ($\approx\SI{20}{Mpc}$) binary black hole event \cite{johnson_prospects_2019,grant_outlook_2023,lasky_detecting_2016}.
Therefore, previous work focuses on detecting memory in the entire population of gravitational-wave events, rather than a single event \cite{lasky_detecting_2016}. 
Searches for memory have been carried out with data from the first \cite{hubner_measuring_2020} and second \cite{hubner_memory_2021} gravitational-wave transient catalogue. 
No evidence of memory was found and Ref.~\cite{hubner_memory_2021} showed that definitive evidence of memory is likely to require an ensemble of $\mathcal{O}(2000)$ gravitational-wave events \cite{hubner_memory_2021, grant_outlook_2023}. 
Proposed future gravitational-wave detectors such as Cosmic Explorer \cite{reitze_cosmic_2019}, the Einstein Telescope \cite{punturo_einstein_2010} and LISA \cite{amaro-seoane_laser_2017} may be able to directly detect memory from a single event \cite{grant_outlook_2023, islo_prospects_2019}.

Theoretical work has shown that gravitational-wave memory is connected to the Bondi-Metzner-Sachs (BMS) symmetry group and Weinberg's soft graviton theorem in quantum field theory \cite{strominger_gravitational_2014, weinberg_infrared_1965}. Each of these three seemingly unrelated concepts represent a corner in the so-called ``infrared triangle'' \cite{strominger_lectures_2018}. These connections may serve as a possible bridge between general relativity and quantum field theory, and can be used to test spacetime symmetries \cite{goncharov_inferring_2023}. These connections to asymptotic symmetries and soft theorems may provide insight to the black hole information paradox \cite{hawking_soft_2016}.

Recent work seeks to test if the memory amplitude is consistent with predictions from general relativity \cite{scargle_detection_2022, goncharov_inferring_2023}.
The premise of these studies is that new physics could produce deviations from general relativity that 
may lead to a different memory amplitude \cite{heisenberg_gravitational_2023}.
Other work explores how the inclusion of memory may help to improve the accuracy of gravitational-wave parameter estimation \cite{mitman_adding_2021,gasparotto_can_2023, xu2024enhancing}. 
Still other publications have discussed the possibility of identifying subsolar-mass compact binary mergers \cite{ebersold_search_2020} and using memory to distinguish between neutron star-black hole binary and binary black hole mergers \cite{tiwari_leveraging_2021}.

In this Paper, we perform a search for gravitational-wave displacement memory using 88 events from LIGO--Virgo--KAGRA's third gravitational-wave transient catalogue (GWTC-3). 
Since we are well short of the $\approx 2000$ events that are expected to be required to detect memory, we view this paper as an ongoing effort to refine the memory detection pipeline and identify potential problems early. In order to search for new physics, we constrain the memory scale factor, which is $A=1$ in general relativity. 
We show that this search is complicated by low-frequency, non-Gaussian noise.
We discuss possible remediation strategies.

The remainder of this Paper is structured as follows. 
In Section \ref{sec:bf}, we describe our search for gravitational-wave memory using data from GWTC-3. 
In Section \ref{sec:posterior}, we describe our search for physics beyond general relativity and our constraints on the memory scale factor $A$, which is expected to be $A=1$ for general relativity. 
In Section \ref{sec:conclusion}, we summarise our results and discuss future research.

\section{Search for gravitational-wave memory with GWTC-3}\label{sec:bf}
We follow the method laid out in Refs.~\cite{hubner_measuring_2020,hubner_memory_2021}, calculating a memory versus no-memory Bayes factor for each event and then adding the log Bayes factors.
However, in this work we use a different waveform.
Reference \cite{hubner_memory_2021} used two waveform approximants: \imrx \cite{garcia-quiros_imrphenomxhm_2020} to cover the extreme mass ratios and \nrsur \cite{Varma_2019} spin precession effects.
In this work, we use a single waveform, \imrxp \cite{pratten_computationally_2021}, which includes extreme mass ratios, spin precession effects, and includes several of the most dominant higher-order modes~\cite{talbot_gravitational-wave_2018}.

The memory component of the gravitational-wave waveform is calculated from the oscillatory component of the waveform by using the \gwmemory \python package \cite{talbot_gravitational-wave_2018}. 
Our analysis consists of two models, a no-memory hypothesis in which our waveform contains only oscillatory wave and a memory hypothesis in which our waveform has both the oscillatory wave and the memory. 
We calculate a memory Bayes factor, which is a ratio of the Bayesian evidence values computed for our two hypotheses:
\begin{equation}\label{BF eq}
    \text{BF}_{\text{mem}} = \frac{\mathcal{Z}_{\text{osc}+\text{mem}}}{\mathcal{Z}_{\text{osc}}} .
\end{equation}

We use data from the Advanced LIGO H1 observatory in Hanford, WA, the LIGO L1 observatory in Livingston, LA \cite{collaboration_advanced_2015} and the Virgo observatory in Italy \cite{acernese_advanced_2014}. 

To calculate the Bayes factors, we use parameter estimation results obtained using the \bilby \cite{Ashton_2019, romero-shaw_bayesian_2020} implementation of \dynesty \cite{Speagle_2020}. 
Where possible, we use results available on the Gravitational-Wave Open Science Centre.\footnote{https://gwosc.org/eventapi/html/GWTC/} 
However, \bilby results are not available for the events GW190725, GW190728, GW190917 and GW190924, and so we generate new results from scratch.
We omit binary neutron star mergers GW170917 and GW190425.
The \imrxp waveform model does not take into account neutron star physics and, at any rate, low-mass binaries produce relatively less memory, making these events expendable for this analysis. 

After performing parameter estimation with the oscillatory waveform, we employ importance sampling on each event in order to reweight the $n$ samples with the oscillatory+memory likelihood \cite{payne_higher_2019}: 
\begin{equation}
    \text{BF}_{\text{mem}} = \frac{1}{n}\sum^{n}_{k=1}w(\theta_k) .
\end{equation}
Here, the weights $w$ are the likelihood ratios comparing our two hypotheses
\begin{equation}
    w_i(\theta_k) =\frac{{\cal L}_{\text{osc}+\text{mem}}(d_i | \theta_k)}{{\cal L}_{\text{osc}}(d_i | \theta_k)} ,
\end{equation}
where $d_i$ is the data for event $i$ and $\theta_k$ are the parameters associated with posterior sample $k$.
We employ a minimum frequency of \SI{20}{Hz}.
The total Bayes factor for GWTC-3 is simply
\begin{equation}
    \text{BF}^{\text{tot}}_{\text{mem}}=\prod^N_{i=1}\text{BF}^i_{\text{mem}} .
\end{equation}
We consider $\ln \text{BF}^{\text{tot}}_{\text{mem}}\geq 8$ to be a detection of memory \cite{lasky_detecting_2016}.

\begin{figure}
    \centering
    \includegraphics[width=0.9\linewidth]{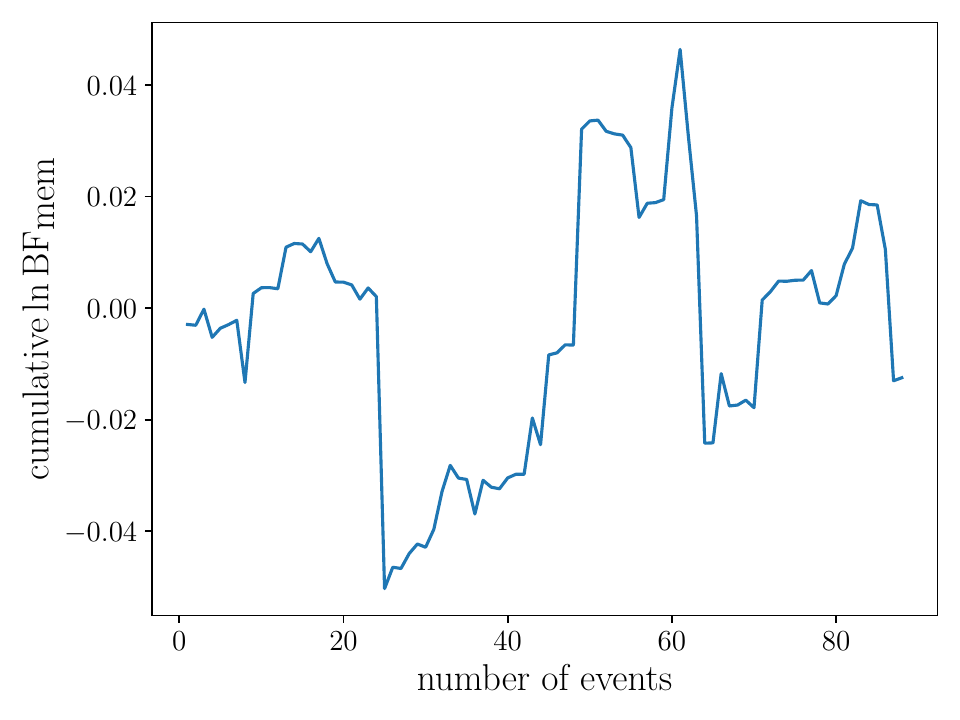}
    \caption{Cumulative natural log Bayes factor $\ln\text{BF}^{\text{tot}}_{\text{mem}}$ as a function of the number of binary black hole mergers. 
    Large positive values indicate support for the existence of memory while large negative values indicate support for the no-memory hypothesis.
    The current data are inadequate to differentiate between these two hypotheses.
    }
    \label{fig:cumulative_lnbf}
\end{figure}

In Fig.~\ref{fig:cumulative_lnbf}, we plot the cumulative Bayes factor as a function of the chronological event number. 
Some events are more informative than others causing comparatively large jumps.
The final value is $\ln \text{BF}^{\text{tot}}_{\text{mem}}=-0.01$, which is too small to favour one hypothesis over the other. 
This is expected as ${\cal O}(2000)$ events are needed before we expect to have the statistical power to distinguish between these two hypotheses \cite{hubner_memory_2021}. The uncertainty from the reweighting method is $<0.01$, much less than the threshold of detection of $8$, see bottom panel of Fig. 1 of \cite{hubner_measuring_2020}.

The Bayes factors for individual events are displayed in Fig. \ref{individual_BF} of the Appendix A, which includes a comparison with previous results and an explanation for differences between this work and Ref.~\cite{hubner_memory_2021}. We omit event GW190424 from our analysis, as it was not deemed a significant event for inclusion in GWTC-2.1 \cite{the_ligo_scientific_collaboration_gwtc-21_2022}, despite being present in previous searches for memory \cite{hubner_memory_2021}.

\section{Search for new physics with non-standard memory}\label{sec:posterior}
In order to search for new physics, and following Refs.~\cite{scargle_detection_2022,goncharov_inferring_2023}, we allow the amplitude of the memory to vary by a multiplicative factor $A$ so that the gravitational-wave signal is
\begin{equation}
    h_{\text{total}} = h_{\text{osc}}(t) + A \, h_{\text{mem}}(t) .
\end{equation}
By construction, general relativity predicts $A=1$.
However, in this framework, we speculate that new physics---perhaps related to Bondi-Metzner-Sachs (BMS) symmetry \cite{bondi_gravitational_1962, sachs_gravitational_1962}---leads to a waveform with $A \neq 1$.
We assume that $A$ is the same for each event and calculate the posterior for $A$ given the events in GWTC-3:
\begin{align}
    p(A | \vec{d}) \propto \pi(A)
    \prod_i^N \int d\theta \, 
    {\cal L}(d_i | A, \theta_i) \pi(\theta_i) .
\end{align}
We take the prior $\pi(A)$ to be uniform on the interval (0,400).

\begin{figure}[h]
    \centering
    \includegraphics[width=0.9\linewidth]{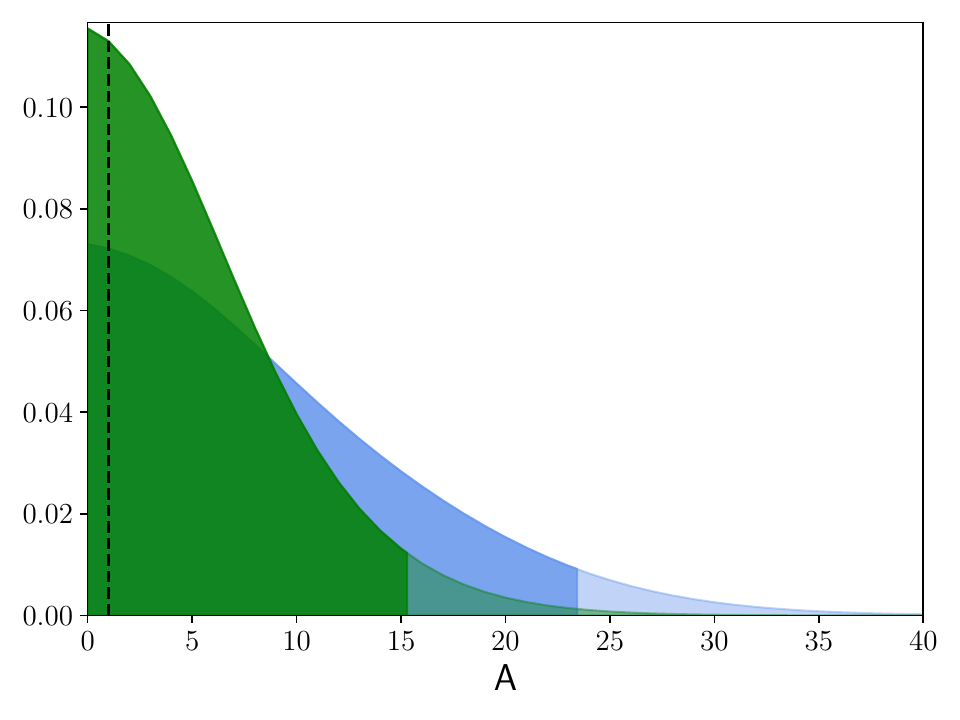}
    \caption{
    The posterior for $A$ given the 88 events in our analysis (blue). 
    The posterior is consistent with $A=1$ (dashed line) predicted by general relativity. 
    The 95\% credible interval (dark blue) yields an upper limit of $A=23$. To better constrain $A$, we remove 3 events most affected by non-stationary noise (GW170104, GW170818, GW200128) from the posterior (green), with an upper limit of $A=15$.
    }
    \label{combined posterior}
\end{figure}

The posterior distribution for $A$ (calculated with all of the events in GWTC-3) is shown in Fig.~\ref{combined posterior}. 
It is consistent with the general relativity prediction of $A=1$.
We set a 95\% upper limit of $A=23$.
Again, we employ a minimum frequency of \SI{20}{Hz}.

\begin{figure}
    \centering
    \includegraphics[width=0.9\linewidth]{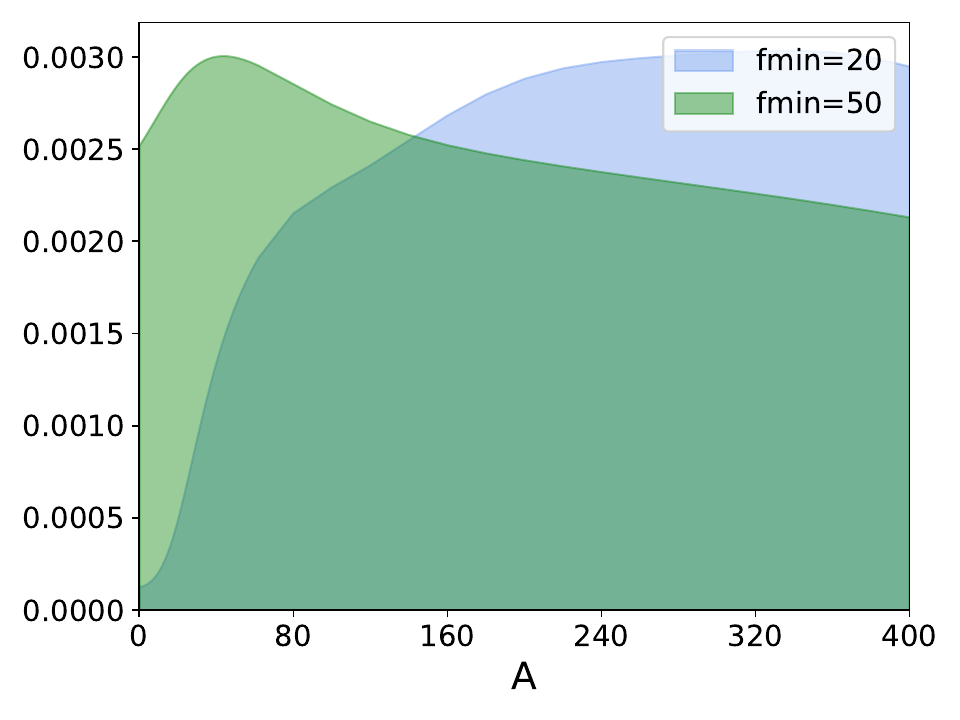}
    \caption{The posterior of $A$ for GW170818. The posterior calculated from a \SI{20}{Hz} (\SI{50}{Hz}) high-passed data is shown in blue (green). The \SI{50}{Hz} high-passed posterior has a stronger support for $A=0$ and $A=1$ and favours a smaller $A$.}
    \label{GW170818_compare_20hz_vs_50hz}
\end{figure}

In the course of carrying out this analysis, we noticed that, for some events, the posterior favours large values of $A$. 
The blue distribution in Fig.~\ref{GW170818_compare_20hz_vs_50hz} shows this behaviour for one such event (GW170818), which favours $A\approx 300$ over $A=1$ with a likelihood ratio of $\approx 20$.
While it is expected that some events will produce posteriors peaked away from $A=1$ due to noise fluctuations, we do not observe such large fluctuations when we repeat the analysis using simulated Gaussian noise.

We confirm that this behaviour is not due to real $A=300$ memory by analyzing ``off-source'' data where no oscillatory gravitational-wave signal has been detected.
We add a gravitational-wave signal to the data with standard $A=1$ memory and calculate the posterior for $A$ with the same procedure.
We observe a similar pattern with $\approx 3/88$ fake events exhibiting large fluctuations away from $A=1$.
We conclude that unmodeled non-Gaussian noise in the LIGO--Virgo data is affecting our posterior for $A$.\footnote{We consider two other hypotheses that might explain why the posterior prefers large values of $A$ for some events. 
First, we do not take into account uncertainty in our estimation of the noise power spectral density \cite{student-t,Biscoveanu}. Second, we do not take into account correlations between frequency bins that arise from so-called finite-duration effects \cite{windows}.
Thus, our likelihood is slightly misspecified due to approximations we make about the noise.
However, we rule out these explanations because we do not see posteriors that favour large values of $A$ when we analyse Gaussian noise with the same slightly misspecified likelihood.}
In hindsight, this is perhaps not surprising as non-stationary noise is known to lurk at low frequencies where memory is most pronounced \cite{abbott_guide_2020, the_ligo_scientific_collaboration_characterization_2016}.

\begin{figure*}[h]
    \centering
    \includegraphics[width=0.9\linewidth]{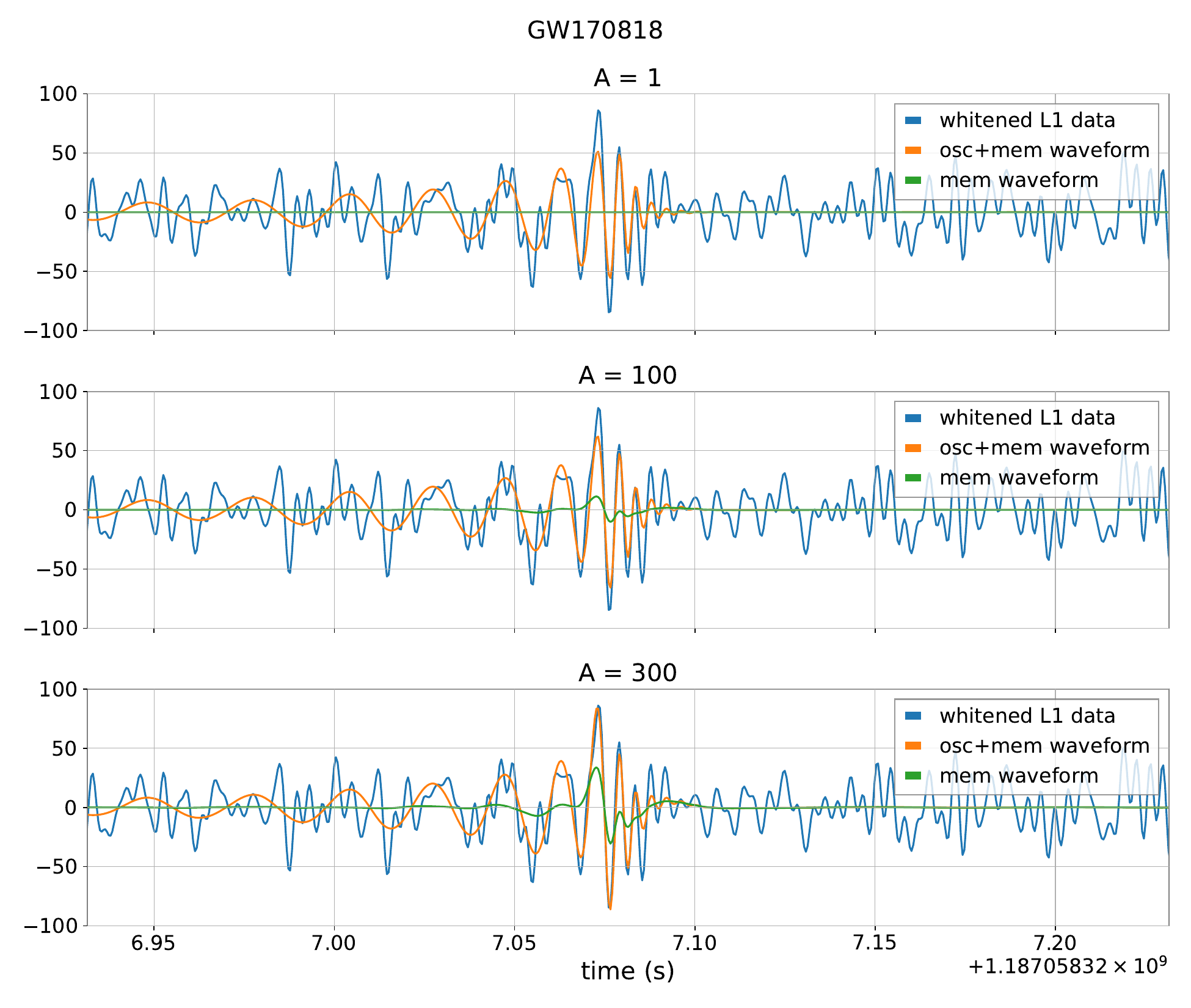}
    \caption{
    A plot of the full (osc+mem) and memory (mem) waveform with the whitened Livingston (L1) data for GW170818. 
    As the $A$ increases, the full waveform fits increasingly well to the whitened data.
    }
    \label{whitened_data_fits}
\end{figure*}

In Fig.~\ref{whitened_data_fits} we provide a visualisation to show how non-Gaussian noise can yield high likelihood values for large values of $A$.
Each panel is a time series of whitened strain.
Blue is Livingston L1 data for GW170818.
The expected memory waveform is in green.
Since we include only frequencies within the LIGO--Virgo observing band (above of minimum frequency of \SI{20}{Hz}), the memory does not induce a DC offset, but instead produces a short-duration wave packet; see also \cite{lasky_detecting_2016}.
The expected oscillatory + memory waveform is in orange.

The top panel shows the expected $A=1$ waveform predicted by general relativity.
For this particular event, the memory is negligible.
However, the oscillatory + memory waveform is not well-matched to the data at the moment of peak strain.
This discrepancy can be plausibly explained as a noise fluctuation since this deviation between orange and blue is not unusual compared to the fluctuations in the noise at late times after the gravitational-wave signal has passed.
The second and third panels show the same plot with $A=100$ and $A=300$.
By increasing the memory amplitude, the waveform better fits the noise fluctuation.
Since the memory signal is so short in duration, this does not spoil the fit with the earlier inspiral phase.
Viewed in the time domain, it is apparent that the short, memory impulses are similar to broadband, low-frequency non-Gaussian noise.

Next, we carried out an investigation in order to identify the frequency band where this non-stationary noise is most pronounced. 
We create a distribution of strain of different frequency bins from 100 random segments of data and fit each distribution with a Gaussian function. 
We expect that all frequency bins contain some non-Gaussian noise. 
However, we find the \SI{20}{Hz} frequency bin to be especially non-Gaussian with reduced chi-squared value of $\bar{\chi}^2=2\times10^5$ whereas the  \SI{50}{Hz} frequency bin $\bar{\chi}^2=1.6$ is more consistent with a Gaussian noise. 
Testing regularly spaced frequency bins, we conclude that the non-Gaussian noise is most pronounced in the band: \SIrange{20}{50}{Hz}. 
This motivates us to see how the results change when we increase the minimum frequency from \SI{20}{Hz} to \SI{50}{Hz}.

In Fig.~\ref{GW170818_compare_20hz_vs_50hz} we compare the posterior for $A$ for GW170818 calculated using $f_\text{min}=\SI{20}{Hz}$ (blue) and $f_\text{min}=\SI{50}{Hz}$ (green). 
The posterior calculated with $f_\text{min}=\SI{50}{Hz}$ is consistent with $A=1$. 
This is consistent with our hypothesis that the $A$ posterior is biased by non-Gaussian noise in the \SIrange{20}{50}{Hz} band. 
We recalculate the posterior distribution by removing the three events that appear to suffer from non-stationary noise the most (GW170104, GW170818 and GW200128) and obtain the green curve in Fig.~\ref{combined posterior}. 
The resulting upper limit on $A$ is reduced to $A=15$. 
Although removing the low-frequency data removes the non-Gaussian noise, it also removes part the memory signal, reducing the optimal SNR by $10-66\%$. 
This reduces our ability to detect memory and to constrain $A$.

We consider various solutions to deal with the non-Gaussian noise at low frequencies.
Instead of throwing out the \SIrange{20}{50}{Hz}, one could model the non-Gaussian noise by developing a more sophisticated likelihood function. 
In this approach the likelihood down-weights the data affected by the non-Gaussian noise as it is less trustworthy. 
The disadvantage of this approach is that the the down-weighting still reduces the sensitivity of the search, though, not as much as removing the low-frequency data entirely.
The best solution is to reduce non-stationary noise with commissioning. 
Investigating this possibility is a goal for future research.

\section{Conclusion}\label{sec:conclusion}

With LIGO-Virgo-KAGRA's fourth observation run already underway and the fifth observation run is planned to start in 2027, the number of gravitational-wave events will greatly increase. 
It is expected we will reach the $\approx 2000$ events needed to detect memory during the fifth observing run. 
Non-Gaussian noise between \SIrange{10}{50}{Hz} needs to be better understood, otherwise the required number of events to detect memory may be larger. 
We suggest mitigating the non-Gaussian noise either through detector commissioning or by developing a model for non-Gaussian noise.

\section{Acknowledgements}\label{sec:acknowledgements}
We would like to thank Moritz Huebner and Colm Talbot for helpful discussions. This research is supported by Australian Research Council (ARC) Centre of Excellence for Gravitational-Wave Discovery CE170100004, Discovery Projects DP220101610 and DP230103088, and LIEF LE210100002. The authors are grateful for computational resources provided by the LIGO Laboratory and supported by National Science Foundation Grants PHY-0757058 and PHY-0823459, and the OzSTAR Australian national facility at Swinburne University of Technology. This material is based upon work supported by NSF's LIGO Laboratory which is a major facility fully funded by the National Science Foundation.

\appendix

\section{Bayes factors for individual events}

\begin{figure*}
    \centering
    \includegraphics[width=0.99\linewidth]{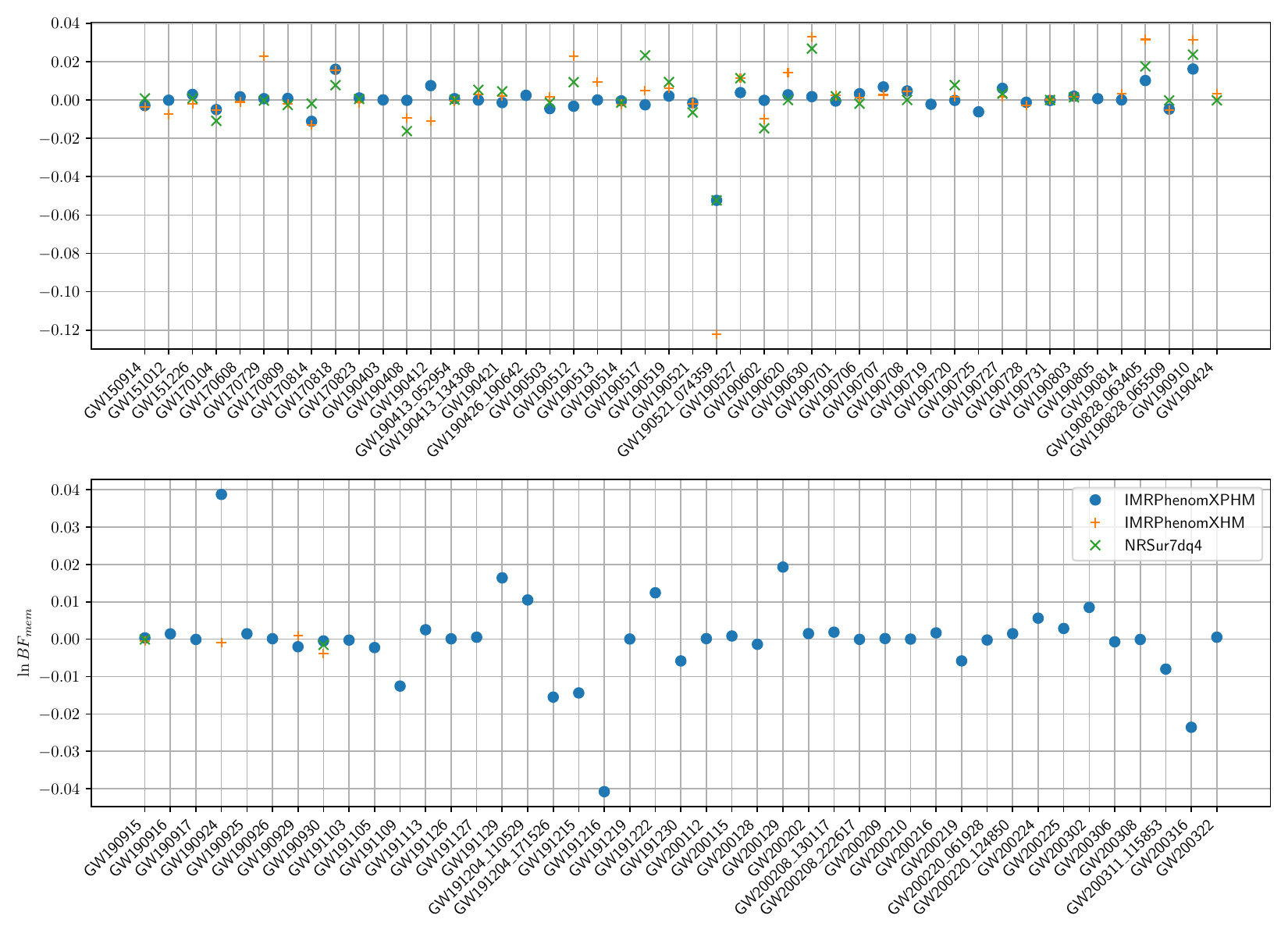}
    \caption{The individual $\ln \text{BF}_{\text{mem}}$ for all 88 binary black hole mergers. Our results computed using \imrxp are shown as blue dots, while the results from \cite{hubner_memory_2021} computed using \imrx and \nrsur are shown as pluses and crosses, respectively. }
    \label{individual_BF}
\end{figure*}

We compare our $\ln \text{BF}_{\text{mem}}$ values for GWTC-1 \cite{abbott_gwtc-1_2019} and GWTC-2 \cite{abbott_gwtc-2_2021} with the previous search of memory in \cite{hubner_memory_2021}, as shown in Figure \ref{individual_BF}. 
For most events, our Bayes factor computed using \imrxp (blue dots in Figure \ref{individual_BF}) is very close to \cite{hubner_memory_2021}, computed using \imrx and \nrsur, hence are in agreement within waveform systematics.

However, there are a few individual events where the Bayes factors are noticeably different. The differences in our results can be attributed to several factors. The largest factor is the changes made in the \gwmemory package between the time of \cite{hubner_memory_2021} and now. The most significant change was switching to an analytic version of the mode amplitudes using Clebsch-Gordan coefficients, aligning memory to the SXS memory prediction. A second factor may be due to systematic difference between the different waveform approximants. To find the difference between waveforms, we run parameter estimation on GW190924, which has the greatest difference in Bayes factor, using the same waveform as \cite{hubner_memory_2021}, \imrx. The difference between our Bayes factor for\imrx and \imrxp is $\approx 0.01$.

\printbibliography

\end{document}